\documentclass[epj]{svjour}

\usepackage{graphicx}
\usepackage{axodraw}
\usepackage{fancyhdr}
\def\makeheadbox{{%  remove EPJC header
 }}
\def\mytitle{My title}
\def\myauthors{My name}
\def\mytype{My type of session}
\def\mysession{My session}
\def\mytitle{Phenomenology of the 1/$N_f$ Expansion} %Put your title here!
\def\myauthors{D.I.Kazakov and G.S.Vartanov}    %Put your name here!
\def\mytype{Contributed Talk}
\def\mysession{Alternatives}
\begin{document}
\title{Phenomenology of the 1/N$_f$ Expansion for Field Theories in Extra Dimensions}
\author{D.I.Kazakov\inst{1,2}
\thanks{\emph{Email:} KazakovD@theor.jinr.ru}%
 \and
 G.S.Vartanov\inst{1,3}% etc
\thanks{\emph{Email:} Vartanov@theor.jinr.ru}%
}                     % Do not remove
\institute{Bogoliubov Laboratory of Theoretical Physics, Joint Institute for Nuclear
Research, Dubna, Russia \and Institute for Theoretical and Experimental Physics, Moscow,
Russia \and University Center, Joint Institute for Nuclear Research, Dubna, Russia}
\date{}
\abstract{ In this paper we review the properties of the 1/$N_f$ expansion in
multidimensional theories. Contrary to the usual perturbative expansion it is
renormalizable and contains only logarithmic divergencies. The price for it is the presence
of ghost states which, however, in certain cases do not contribute to physical amplitudes.
In this case the theory is unitary and one can calculate the cross-sections. As an example
we consider the differential cross section of elastic $eq \rightarrow eq$ scattering in
$D=7,11,...$-dimensional world. We look also for the unification of the gauge couplings in
multidimensional Standard Model and  its SUSY extension which takes place at energies lower
than in 4 dimensions.
 % end of PACS codes
} %end of abstract
\maketitle
%DO NOT REMOVE THIS LINE
%

\section{Introduction}\label{intro}
Theories in extra dimensions are  one of the best candidates for a new physics beyond the
Standard Model~\cite{Original,review}. The drawback of these theories is
non-renormalizabi\-li\-ty since the coupling constant has negative dimension and
perturbative expansion is seek of ultraviolet divergencies.  Thus, one can deal with such
theories as effective theories~\cite{ratt} valid up to some scale and consider them at
the tree level in hope that some higher energy theory like a string theory will  cure all
the UV problems.

Recently we have shown that using the $1/N_f$ expansion~\cite{KV} the theory in
multi-dimensional world is renormalizable and has only logarithmic divergencies contrary to
the usual perturbative expansion.  In this approach the propagators of the gauge fields
acquire the  improved UV behaviour $(p^2)^{1-D/2}$ which has better convergence in loop
integrals. This behaviour corresponds to conformal fixed points and leads to logarithmic
divergences in any space time dimension greater than 4. For $D$ odd one has a non-integer
power that manifests the idea of unparticle physics proposed recently by
Georgi~\cite{Georgi}.

Effectively the $1/N_f$ expansion leads to the higher derivative theories which suffer from
the presence of the ghost states widely discussed in the
literature~\cite{Nexpansion},~\cite{unitarity}. We argued in~\cite{KV} that in some cases
these ghost states might be irrelevant since they do not contribute to the physical
amplitudes. We present these arguments below. Provided this is true one can consider such
amplitudes and calculate the cross-sections to compare them with the ones in 4 dimensional
theories. We consider one of such examples.

Several years ago it was suggested that in extra dimensions the running of the couplings is
power-like which leads to the early unification~\cite{powerbehav}. That was based on the
Kaluza-Klein picture of extra dimensions and a special cut-off procedure of higher modes
due to the non-renormalizability of the underlying theory. Contrary, in the case of the
$1/N_f$ expansion in extra dimensions one has the renormalizable theory with logarithmic
running, however, faster than in 4 dimensions. It leads to the approximate unification both
in the Standard Model and in the MSSM.

\section{The $1/N_f$ Expansion for Extra-Dimensional Theories}\label{1Nexpansion}

Here we briefly discuss the main features of the $1/N_f$ expansion in multidimensional
space-time~\cite{KV}. As an example we consider the usual QED with $N_f$ fermion fields
in $D$ dimensions, where $D$ takes an arbitrary odd value. The Lagrangian looks like
{\small
\begin{equation}\label{l}
  {\cal L} =
    - \frac 14 F_{\mu\nu}^2 -
    \frac {1}{2\alpha} (\partial_{\mu}A_{\mu})^2 + \bar{\psi_i} (i\hat{\partial}-m)
  \psi_i
  + \frac{e}{\sqrt{N_f}} \bar{\psi_i} \hat{A} \psi_i,
\end{equation}}
where $F_{\mu\nu} \ = \ \partial_\mu A_{\nu} -
\partial_\nu A_{\mu}$.

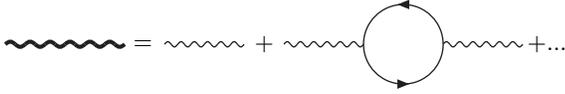
\begin{figure}
\begin{picture}(215,70)(-10,-35) \SetWidth{1.5}
\Photon(0,0)(45,0){1}{6} \SetWidth{0.5} \Text(52,0)[]{=} \Photon(60,0)(90,0){1}{6}
\Text(98,0)[]{+} \Photon(105,0)(135,0){1}{6} \ArrowArc(150,0)(15,0,180)
\ArrowArc(150,0)(15,180,360) \Photon(165,0)(195,0){1}{6}
 \Text(205,0)[]{+...}
\end{picture}
\caption{The chain of diagrams giving a contribution to the photon propagator in the zeroth
order of the $1/N_f$ expansion}\label{photonPict}
\end{figure}

According to the general strategy of the $1/N_f$ expansion~\cite{Nexpansion} one first
calculates the photon propagator in the leading order of $1/N_f$. Due to transversality of
the polarization operator it is useful to  apply the Landau gauge. Then in the leading
order one has the following sequence of bubbles (see Fig.(\ref{photonPict}))
 summed up into a geometrical progression. The resulting photon propagator
takes the form
\begin{equation}\label{photonprop}
  D_{\mu\nu}(p) = -\frac{i}{p^2}\left(g^{\mu\nu}-\frac{p^\mu p^\nu}{p^2}\right)
  \frac{1}{1 + e^2 f(D)(-p^2)^{D/2-2}},
\end{equation}
where
$$f(D)=\frac{\Gamma^2(D/2)\Gamma(2-D/2)}{2^{D-[D/2]-1}\Gamma(D)\pi^{D/2}}$$
and we put $m=0$ for simplicity.

It useful now to change the normalization of the gauge field $A_\mu\to A_\mu/e$ and
introduce the dimensionless coupling $h$ associated with the triple vertex for the
interaction of fermions with the gauge field~\cite{KV}. The new coupling $h$ is needed
for the multiplicative renormalizability of the theory and provides the validity of the
renormalization group pole equations~\cite{Hooft}. After this the effective Lagrangian
takes the form {\footnotesize
\begin{eqnarray}\label{eff}   {\cal L}_{eff} &=& - \frac 14
F_{\mu\nu}\left( \frac{1}{e^2}
  + f(D)(\partial^2)^{D/2-2} (1+h)
  \right) F_{\mu\nu} \\
&&  - \frac {1}{2\alpha e^2} (\partial_{\mu}A_{\mu})^2  + \bar{\psi_i} (i\hat{\partial}-m)
  \psi_i + \frac{\sqrt{h}}{\sqrt{N_f}} \bar{\psi_i} \hat{A} \psi_i.
\nonumber
\end{eqnarray}
} In the non-Abelian case in addition one has the triple and quartic self-interaction of
the gauge fields. These vertices, which are suppressed by $1/\sqrt{N_f}$ and $1/N_f$,
respectively, obtain loop corrections of the same order in $1/N_f$. This leads to extra
terms in effective Lagrangian. For more details see~\cite{KV}.

The Lagrangian (\ref{eff}) is written in multi-dimensional world. At high energies
($E>M_c$, where $M_c$ is the scale of extra dimensions) the higher-derivative term
dominates, while at low energies ($E<M_c$) one has the usual Maxwell term, thus
establishing the correspondence with the classical theory. To find connection to the
four-dimensional world one has to assume some kind of compactification or localization. For
example, within the brane-world scenario one may take the fermion fields localized on 4-dim
brane and integrate over the extra dimensions. Then one gets the following relation between
the D-dimensional and four dimensional couplings
$$ e_{(D)}=e_{(4)}/V_d,\ \ \ e_{(4)}\equiv\sqrt{h},$$
where $V_d\sim M_c^{d}$ is the volume of compact extra dimensions or localization volume.
The new dimensionless coupling $h$ enters into the gauge transformation and plays the role
of the gauge charge.

As follows from eq.(\ref{eff}) one has the modified Feynman rules with the photon
propagator that decreases in the Euclidean region like $1/(p^2)^{D/2-1}$, thus improving
the UV behaviour in a theory. The only divergent graphs are those of the fermion propagator
and the triple vertex. They are both logarithmically divergent for any odd D. The photon
propagator is genuinely finite and may contain divergencies only in subgraphs. Therefore,
due to the Ward identities in QED, one has vanishing beta-function and the coupling $h$ is
not running. In the non-Abelian case this is not true and the beta-function  is given by
{\footnotesize
\begin{equation}\label{betaFunc}
\beta(h)= \frac{-\Gamma(D)(D-1)C_A  \frac{\displaystyle h^2}{\displaystyle (1+h)^2} }{2^{[D/2]+2} \Gamma(2-D/2) \Gamma(D/2+1)
\Gamma^2(D/2)N_fT}.
\end{equation}
} One can see that $\beta(h)>0$ for $D=5,9,...$, $\beta(h)<0$ for $D=7,11,...$ and then
alternates with $D$. Hence, one has UV asymptotic freedom for dimensions $D=7,11,...$

Consider now the analytical properties of the propagator (\ref{photonprop}) and related
problem of unitarity. Besides the cut starting from $4m^2$ it has poles in the complex
$p^2$ plane. Hence, knowing the analytical structure, one can write down the
K\"allen-Lehmann representation. It has two contributions: one comes from the pole
structure and the other one is the continuous spectrum. Depending on a sign of $f(D)$
there are two possibilities: either one has a pole at the real axis and (possibly) pairs
of complex conjugated poles ($f(D)<0$, D=5,9,...) or one has only pairs of complex
conjugated poles ($f(D)>0$, D=7,11,...) and all the rest appears at the second Riemann
sheet. In~\cite{KV} it was shown that the continuous spectrum has a positive spectral
density and corresponds to the production of pairs of fermions. These states are present
in the original spectrum and cause no problem with unitarity. One can show that all the
cuts imposed on diagrams when applying Cutkosky rules in any order of perturbation theory
lead to the usual asymptotic states on mass shell and no new states appear.

\begin{figure}
\begin{picture}(240,130)(0,-70)

\SetWidth{0.5} \Line(0,0)(50,0) \Line(50,-50)(50,50) \SetWidth{2} \Line(50,0)(100,0)
\SetWidth{0.5} \Line(130,0)(180,0) \SetWidth{2} \Line(180,0)(230,0) \SetWidth{0.5}
\Line(180,-50)(180,50) \Vertex(25,0){2} \Vertex(200,-20){2} \Vertex(200,20){2}
\Text(10,40)[]{D=5} \Text(140,40)[]{D=7} \Text(95,-10)[]{Re} \Text(60,40)[]{Im}
\Text(225,-10)[]{Re} \Text(190,40)[]{Im}
\end{picture}
\caption{The analytical structure of the gauge field propagator}\label{Ghost}
\end{figure}
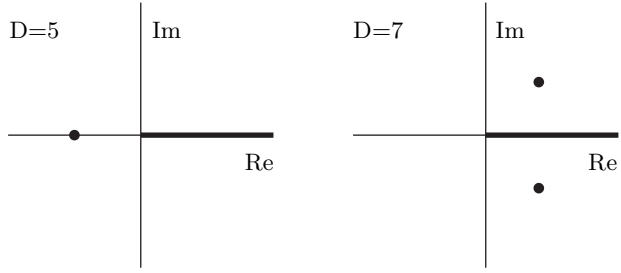

The problem comes with the poles. One can see that the pole terms come with negative sign
and, therefore, correspond to the ghost states. For $D=5$ one has only one pole at the
positive real semiaxis while for $D=7$ one has a pair of complex conjugated poles, as shown
in Fig.~\ref{Ghost}.

The presence of those ghost states is the drawback of a theory. They signal of
instability of the vacuum state. So one has either to try to get rid of the ghost poles
or to make sure that they do not give a contribution to the physical amplitudes. It was
shown in~\cite{KV} that the contribution of those states to the physical amplitudes is
canceled in $D=7,11,...$ due to the complex conjugation of the poles in
(\ref{photonprop}) and they do give contribution to the physical amplitudes in
$D=5,9,...$

\section{The Cross Section of Elastic Scattering}\label{XSection}

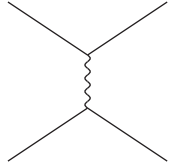
\begin{figure}
\begin{picture}(100,80)(-90,0)
\Line(0,20)(30,40) \Line(30,40)(60,20) \Photon(30,40)(30,60){1}{4} \Line(0,80)(30,60)
\Line(30,60)(60,80)
\end{picture}
\caption{The $e^-q \rightarrow e^-q$ scattering process} \label{fig1}
\end{figure}

To find out the phenomenological consequences of extra-dimensional theories within our
approach we calculate the cross section of elastic  $e^-q$ scattering: $e^-q \rightarrow
e^-q$. In the Standard Model for this process one has only one diagram in $t$-channel
(see Fig.(\ref{fig1}))
%The amplitude
%of this diagram for SM in four dimensions is
%\begin{equation}\label{ampSM}
%  Amp\ \sim\   \bar u(p_1)\gamma^\mu u(p_2)\
%  \frac{e^2}{t} \
%  \bar u(p_3)\gamma^\mu u(p_4),
%\end{equation}
that gives  the differential cross section~\cite{Peskin}
\begin{equation}\label{diffsec}
\frac{d\sigma_{SM}}{dt} \ = \ \frac{2\pi\alpha^2Q_f^2}{s^2} \left( \frac{s^2+u^2}{t^2}
\right).
\end{equation}

To calculate the same cross section in multi-dimen\-sio\-nal world we assume the brane
world scenario~\cite{local} and consider the scattering of particles that are localized on
4-dim brane while the photon field propagates in the bulk and has the propagator given by
(\ref{photonprop}). The interaction term on the brane now looks like
$$\frac{\sqrt{h}}{\sqrt{N_f}}\int d^4xd^dy \bar \psi(x)\gamma^\mu A_\mu^a(x,y)\psi(x)
\delta^d(y).$$

Given this setup, to get the cross section one just has  to replace the photon propagator
in (\ref{diffsec}) by a modified one and to integrate over the extra dimensions: {\small
%\begin{eqnarray}\label{amp}
$$  \frac 1t \ \Rightarrow \
  \frac{(-)^{d/2}}{(2\pi)^d}\int \frac{d^dp}{(t-p^2)\left( 1/e^2_{(D)} +
N_ff(D)(p^2-t)^{d/2}\right)}$$
%\end{eqnarray}
}

To compare the cross section in multidimensional world with the one in the Standard Model
we assume that the  extra dimensions appear at some scale $M_c$: below this scale on has
the usual 4-dim cross section and above this scale the full multi-dimensional theory is
applied. So one has to merge the two  cross sections at compactification scale $M_c$.
Working in the center of mass frame and taking $\theta=\pi/2$ one gets $t=-s/2, u=-s/2$.
Requiring that
\begin{equation}\label{3}
\frac{d\sigma_{SM}}{dt} |_{s=M_c^2} \ = \ \frac{d\sigma_{D}}{dt} |_{s=M_c^2 },
\end{equation}
one gets the equation for $e^2_{(D)}$.  Taking for example $D=7,11$  and $N_f=3$ we find
numerically $1/e^2_{(7)} \simeq 0.2988M_c^3, 1/e^2_{(11)} \simeq 312.96M_c^7$. Finally we
obtain the following expression for the differential cross section in $D=4+d$ dimensions in
the c.m.f  {\small
\begin{eqnarray}\label{4}
\frac{d\sigma_{D}}{dt} && =  \frac{5\pi e^2Q_f^2}{2(2\pi)^{2d}} \\ \nonumber &&
\hspace*{-1cm}\left( \int \frac{d^dp}{(s/2+p^2)\left( X M_c^d +
3f(D)(s/2+p^2)^{d/2}\right)} \right)^2,
\end{eqnarray}
}where $X$ is found from (\ref{3}) and it depends on the number of dimensions $D$.

\begin{figure}
\includegraphics[width=200pt,height=150pt]{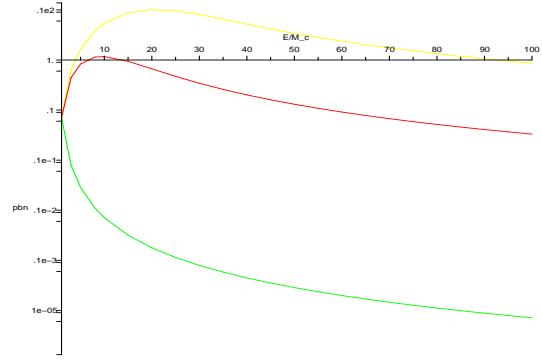}
\caption{The differential cross section for $eq \rightarrow eq$ scattering}\label{cross1N}
\end{figure}

As an illustration we show in Fig.\ref{cross1N} the resulting cross section in the case of
$D=7,11$ for $\theta=\pi/2$. The compactification scale $M_c$ is taken to be equal to 1
TeV. For comparison  we also show the 4-dimensional cross section (\ref{diffsec}).

One can see, that above the compactification scale the cross sections differ essentially,
however, asymptotically the exhibit the same behaviour decreasing like $1/s$ contrary to
Kaluza-Klein approach and in agreement with the fixed point behaviour~\cite{TMP}.

\section{The Gauge Coupling Unification}\label{Unification}

Here we would like to apply the $1/N_f$ expansion to the problem of the gauge coupling
unification. As it is well known, the gauge coupling unification does not take place in
the Standard Model but can be easily achieved in the MSSM. It occurs at very high energy
scale of the order of $10^{16}$ GeV which is far beyond direct search. At the same time
in~\cite{powerbehav} it was suggested that one can get early unification in theories in
extra dimensions due to the power law running of the couplings.

Our approach to extra-dimensional theories based on the $1/N_f$ expansion allows us to look
at this problem differently. We get the logariphmic running but with slightly different
beta functions: for the $U(1)$ coupling constant we have the zero beta function that means
that the $U(1)$ coupling does not run and for $SU(2)$ and $SU(3)$ couplings we have the
beta function given by eq.(\ref{betaFunc}). Substituting the proper Casimir operators and
taking the number of families $N_F=3$ one can solve the RG equations numerically and check
the unification.

We apply the following procedure: Below the unification scale we take the usual RG
equations. Then we use the values of the couplings at the compactification scale as
initial conditions for RG equations in extra dimensions and run the dimensionless
couplings $\alpha_i=h_i/4\pi$ according to new equations. The scale of compactification
is varied to get better unification.

\begin{figure}
\includegraphics[width=150pt,height=125pt]{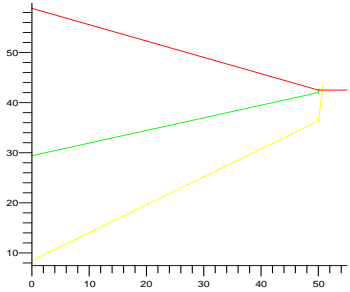}
\caption{The running of the coupling constants in the SM} \label{fig4}
\end{figure}

The resulting curves for the inverse couplings in the case of the Standard Model are shown
in Fig.\ref{fig4} for $D=7$. We would like to stress here that we do not have exact
unification at one point but rather the triangle of the unification though much smaller
than in the usual 4-dim SM. The compactification scale happens to be of the order of $M_c
\sim 7 \cdot 10^{12}$ GeV.

\begin{figure}
\includegraphics[width=110pt,height=120pt]{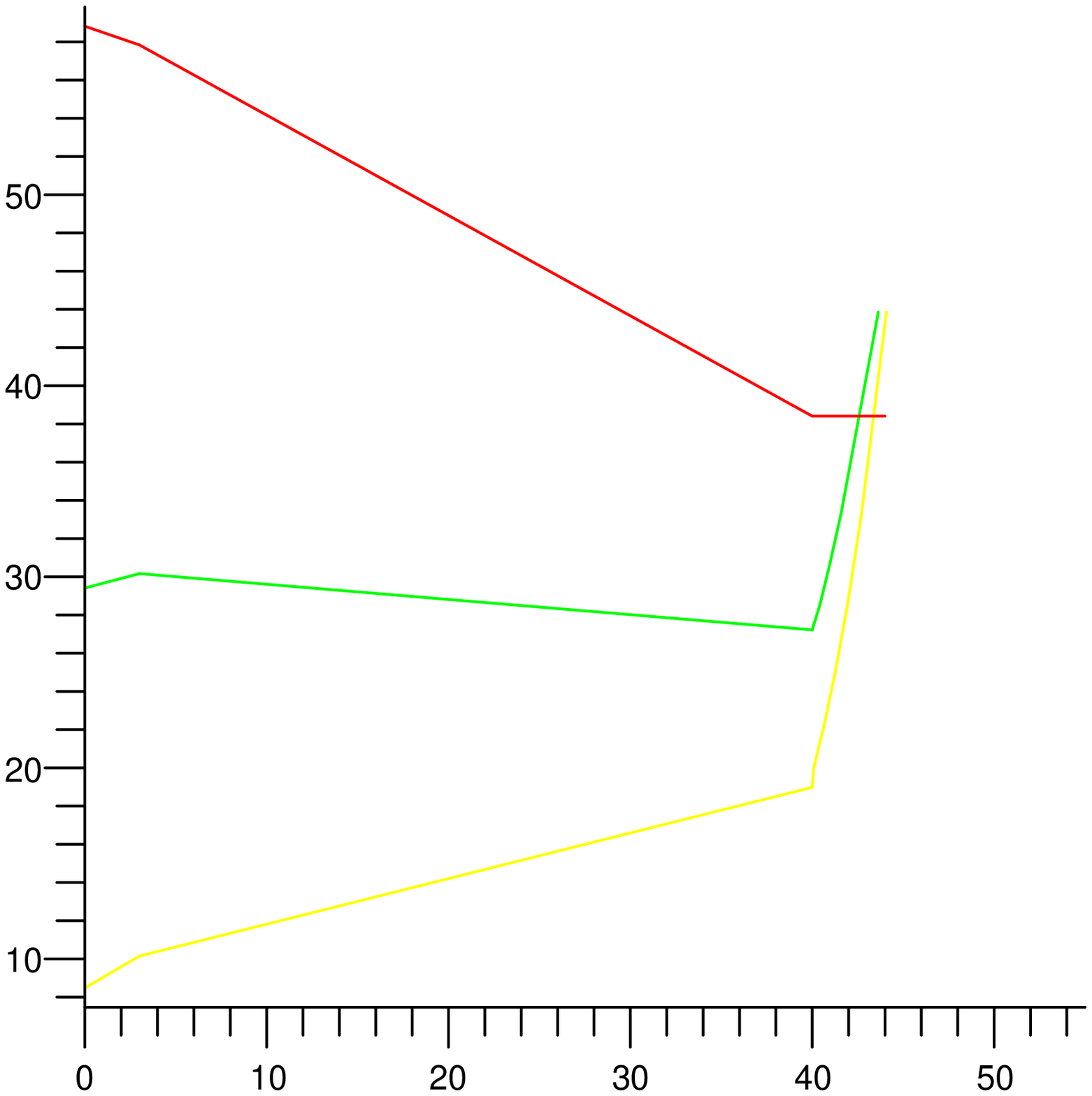}
\includegraphics[width=110pt,height=120pt]{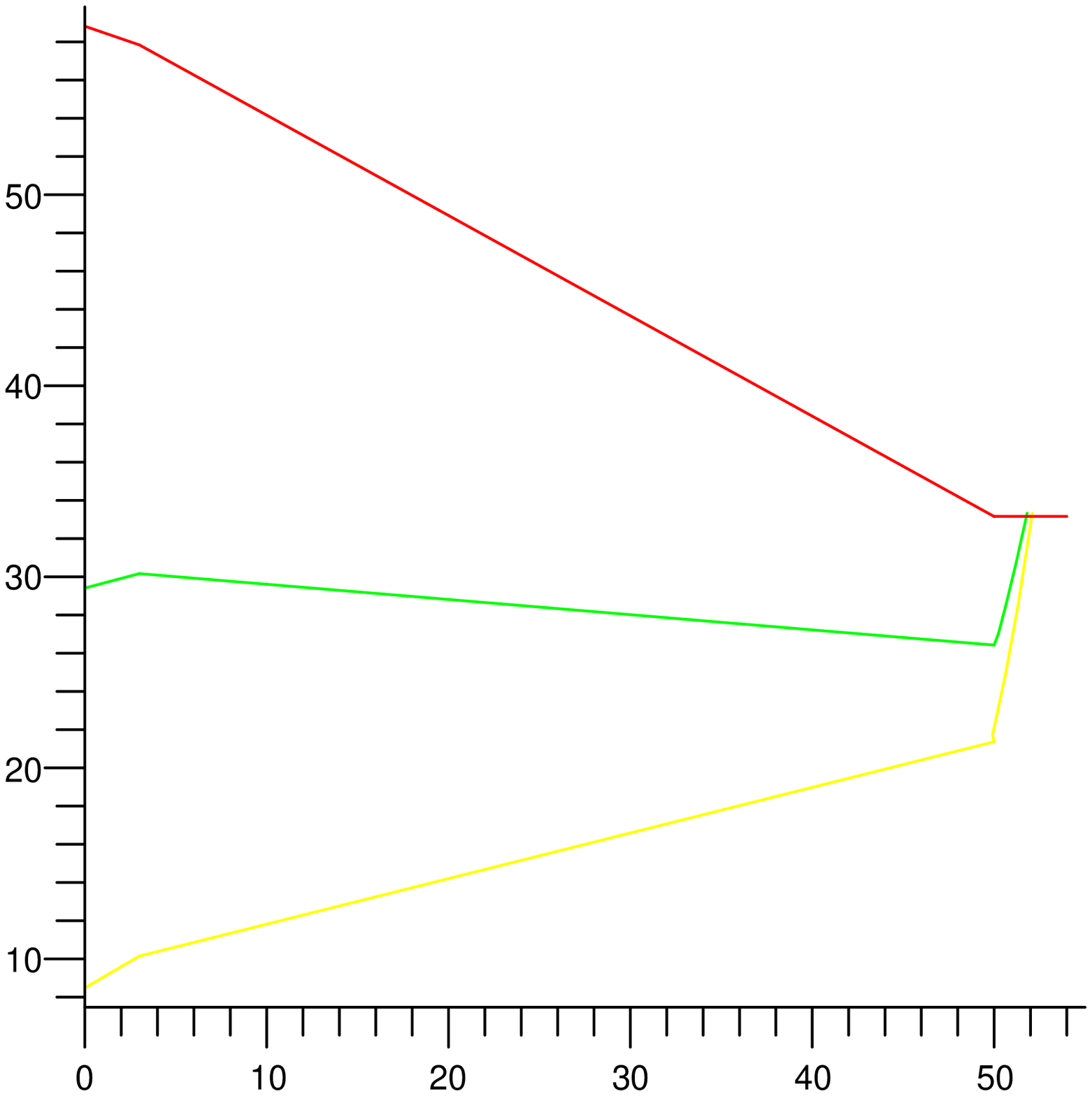}
\caption{The running of the coupling constants in the MSSM}\label{fig5}
\end{figure}

In the MSSM case one has to add the contributions of $N_f$ superpartners, however, the
picture does not change much: qualitatively, unlike the 4-dim case,  one has the same
pattern. In Fig.\ref{fig5} for $D=7$ we show the plots for two possible choices of the
compactification scale: on the left plot we take $M_c \sim 4 \cdot 10^{10}$ GeV and on
the right plot $M_c \sim 7 \cdot 10^{12}$ GeV. In both the cases the SUSY scale is taken
to be the same $M_{SuSy} \sim 500$ GeV.
Again one can get approximate unification, the better the larger the compactification
scale. From this point of view we do not see here much difference between the SM and the
MSSM.

\section{Conclusion}\label{Concl}

We conclude that within the $1/N_f$ approach the multi-dimensional theory can be made
meaningful, renormalizable and manageable. And though it necessarily contains the ghost
states in case when they are complex conjugated their contribution to physical amplitudes
cancels and the theory seems to be unitary.

In such a theory one can calculate the cross sections of elementary processes assuming
certain compactification (localization) scenario. These cross sections differ from those
of the Standard Model in 4 dimensions but asymptotically have the na\"ive scaling
behaviour and, hence, do not violate unitarity.

The running of dimensionless couplings in such theories remains logarithmic and allows one
to get approximate unification of the gauge couplings both in the Standard Model and in the
MSSM, though the unification scale happens to be rather high, yet lower than in the usual
MSSM.

\section*{Acknowledgements}

Financial support from RFBR grant \# 05-02-17603, DFG grant 436 RUS 113/626/0-1 and the
Heisenberg- Landau Program is kindly acknowledged. We want to thank the organizers of the
SuSy'07 conference for warm hospitality.

\end{document}